\documentclass[preprint, aps, amsmath, amssymb, groupedaddress]{revtex4}
\usepackage{graphicx}
\usepackage{amsmath}
\usepackage{color}
\begin{document}

\title{Electrical domain writing and nanoscale potential modulation on LaVO$_3$/SrTiO$_3$}

\author{Mohammad Balal$^1$}
\author{Shilpa Sanwlani$^1$}
\author{Neha Wadehra$^2$}
\author{Suvankar Chakraverty$^2$}
\author{Goutam Sheet$^1$}

\affiliation{$^1$Department of Physical Sciences,
Indian Institute of Science Education and Research Mohali,
Sector 81, S. A. S. Nagar, Manauli, PO: 140306, India}
\affiliation{$^2$Institute of Nano Science and Technology, Habitat Centre, Sector - 64, Phase X, Mohali
Punjab, India - 160062}

\begin{abstract}

The high-mobility 2DEGs formed at the interfaces between certain insulating perovskite oxides have been known to be a novel playground of exotic physical orders like, superconductivity and ferromagnetism and their inter-coupling. There have been efforts to accomplish even more exotic properties at such interfaces of oxide heterostructures through nano-structuring of the surface. In this paper we report writing and erasing charge domains on such an oxide heterostructure LaVO$_3$/SrTiO$_3$ using a conductive AFM cantilever. We have patterned these domains in a periodic fashion in order to create artificial lattices on the surface. Through kelvin probe microscopy, electrostatic force microscopy and conductivity mapping of such artificial lattices we found that the domains not only trap charge carriers but also develop a controllable potential landscape on the surface which coincides with a modulation of local electrical conductivity. The ability to pattern such nanostructures reversibly offers unprecedented opportunities of realizing ultra-high storage density devices in high mobility oxide heterostructures.

\end{abstract}

\maketitle

The interfacial quasi 2 dimensional electron gas (2DEG) formed at the interfaces between two oxide band insulators SrTiO$_3$(STO) and LaAlO$_3$(LAO) has spurted tremendous amount of research in contemporary condensed matter physics\cite{Ohtomo, Cen, Irvin, Li, Popok}. Such systems have shown wide range of exotic physics including the coexistence of gate-tunable superconductivity and magnetism\cite{Thiel, Chen, Brinkman, Kalisky, Bi, Reyren, Klimin, Dikin}. The range of studies have extended beyond the experiments and theories probing the fundamental properties of such 2DEGs. In terms of application potential, the conducting 2DEGs provide the advantage of high concentration of charge carriers without the traditional semiconductor doping mechanism, which would otherwise subsequently reduce the carrier mobility due to scattering with dopants. Therefore a much higher carrier mobility can be expected in the aforementioned oxide 2DEGs which, if realized, would ultimately translate to faster and better devices.  In case of 2DEGs at the oxide bilayers, instead of LAO, which is a band insulator, Mott insulators have also been used where interfacial 2DEGs have been realized\cite{Hotta, He, Choi, YHot}. In such systems, due to strong electronic correlations at the interfacial 2DEGs might give rise to more exotic physics and applicattion potentials. In this paper we focus on one such system involving a band insulator STO and a Mott insulator LVO, where a conducting 2DEG is known to form at the interface when the thickness of LVO exceeds a critical thickness d$_c$=$4$ unit cells\cite{Hotta}. We report a dynamic method for creating artificial patterns on LVO/STO. Such patterns can be written and erased controllably with external electric fields. In this method, a conductive AFM tip is used to induce extremely small electrically active charge domains on the surface. The versatility of creating such nano-domains by this method is that the features can be written and erased infinite number of times with nano-meter scale precision. We have performed a number of scanning probe-based imaging techniques to fully characterize the physical properties and the functionalities of such nano-domains as described below.

Crystalline LVO films were grown on TiO$_2$-terminated STO (001) single crystals using pulsed laser deposition (PLD). For TiO$_2$-termination, the STO substrates were annealed in air at 900$^\circ$C for 2 hours. Steps and terraces along with agglomerates of SrO were formed over the surface of STO crystals during the annealing process. This was followed by etching with hot (60$^\circ$C)deionized (DI) water to remove the SrO agglomerates formed. The quality of the steps and terrace like structure was further improved by annealing the substrates again at 900$^\circ$C for 2 hours.
\begin{figure}
\centering
\includegraphics[width=0.70\textwidth]{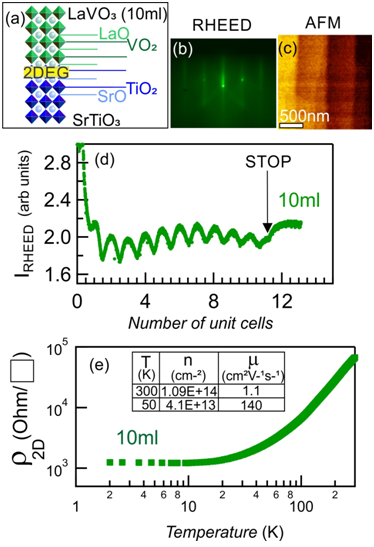}
\caption{(a) Schematic of 2-dimensional electron gas formation at the interface of LVO-STO heterostructure. (b) RHEED pattern of 10ML LVO grown over TiO$_2$-terminated STO substrate with orientation (001). (c) (2$\mu$m x 2$\mu$m) AFM image of LVO grown over STO. Step and terrace like structure on the surface gives clear indication of TiO2 termination. (d) RHEED intensity oscillations during the growth of LVO, as a function of number of unit cells. (e) Temperature dependent two-dimensional resistivity ($\rho_{2D}$)for 10ML sample. Inset table shows the charge carrier density and mobility of the sample at 300K and 50K.}
	\label{Figure 1}
\end{figure}

Crystalline LVO-STO samples were prepared for two different thickness of 4 monolayers (ML) and 10 monolayers of LVO using the same ceramic target. We have chosen these two different thickness of LVO, taking into account the critical thickness (d$_c$=$4$ unit cells) \cite{Hotta} above which the 2DEG formation is achieved. The sample with 4ML of LVO ia at the verge of Metal-Insulator (MI) transition where resistance increases sharply with decreasing temperature. On the other hand 10ML LVO sample was conducting down to 2K. The schematic of the heterostructure grown is shown in Figure 1(a). These depositions were done at substrate temperature of 600$^\circ$C and oxygen partial pressure of 1$\times 10^{-6}$ Torr. The thickness of the films was monitored using reflection high-energy electron diffraction (RHEED) technique. The RHEED intensity oscillations of the specular spot, for 10ML sample, as a function of number of unit cells are shown in Figure 1(d). Figure 1(b) and (c) show the RHEED pattern and AFM image of the 10ML LVO film grown over STO substrate. The sample was found to be conducting down to 2K as shown by the temperature dependent two-dimensional resistivity plot in figure 1(e). The charge carrier densities, for this sample, calculated from the hall measurements done at 300 K and 50 K were 1.09$\times 10^{14} cm^{-2}$ and 4.1$\times 10^{13} cm^{-2}$ respectively. The corresponding mobilities were found to be 1.1$cm^2V^{-1}s^{-1}$ and 140$cm^2V^{-1}s^{-1}$.

\begin{figure}[t]
\centering
\includegraphics[width=0.70\textwidth]{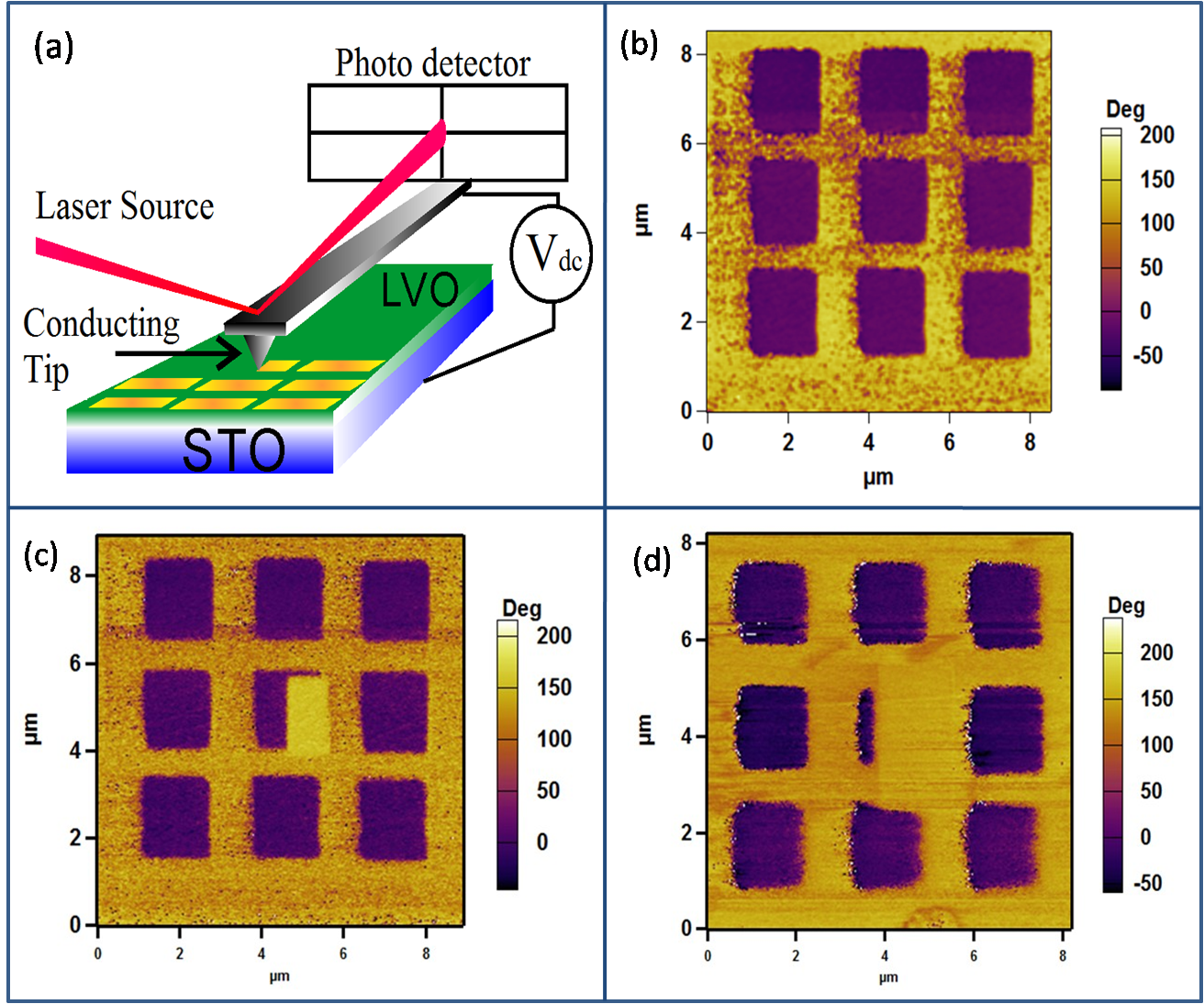}
\caption{(a) Schematic of the contact mode AFM lithographic technique for creating artificial lattice structures (pioneered by the PI). (b) Phase image of the square patterned domains written at room temperature for 4ML LVO-STO by applying +40V (c) Phase image of the domain while applying -40V at the centre of the pattern for 4ML LVO-STO (d) for 10ML LVO-STO. }
	\label{Figure 2}
\end{figure}

\begin{figure}[t]
\centering
\includegraphics[width=0.70\textwidth]{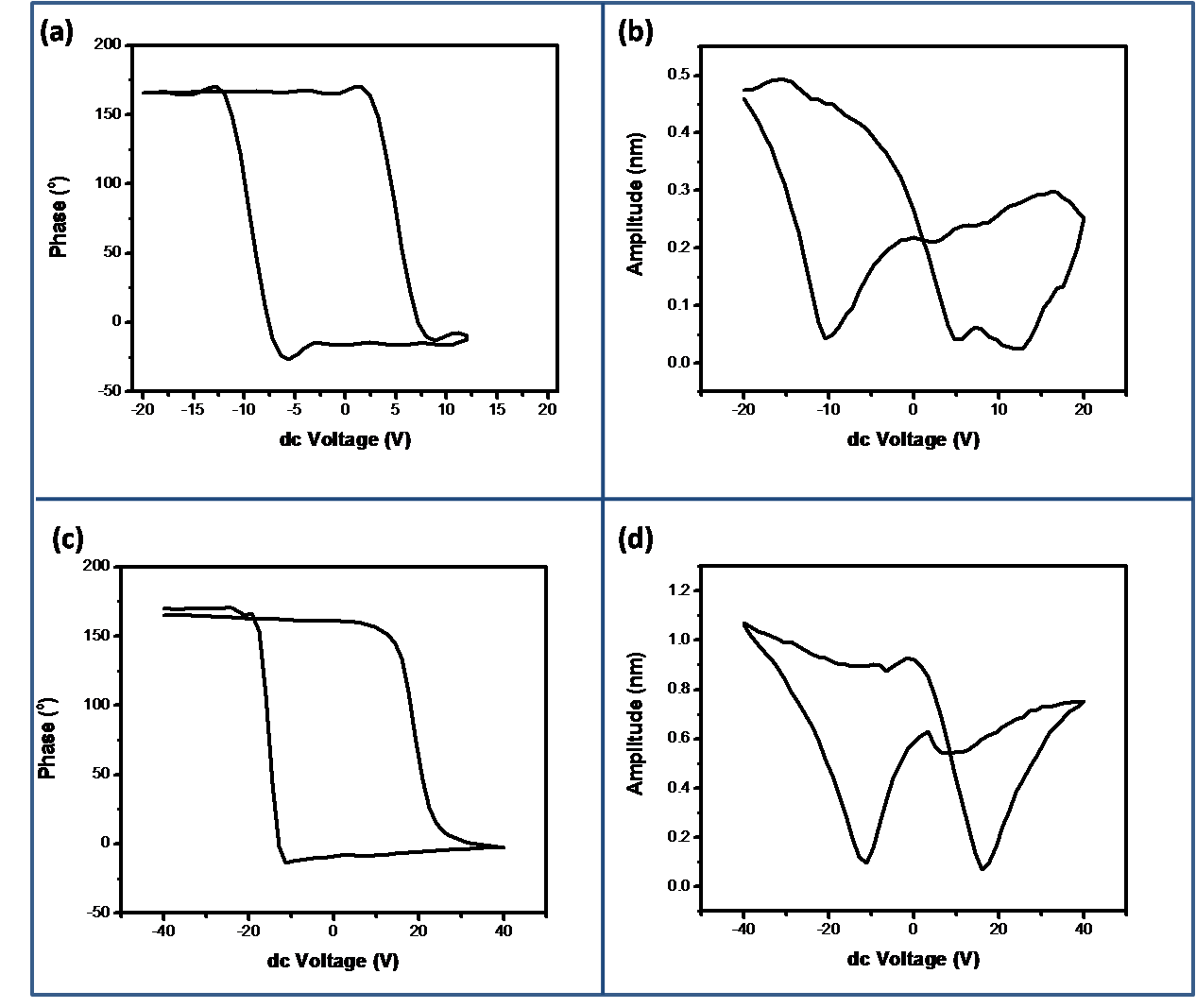}
\caption{ (a) PFM Phase hysteresis and (b) butterfly loop for 4ML LVO on STO substrate at room temperature in the "off" state (c) PFM Phase hysteresis and (d) butterfly loop for 10ML LVO on STO substrate at room temperature in the "off" state }
\label{Figure 3}
\end{figure}

We have performed cantilever based electrical lithography to write electrical domains on LVO using a conducting cantilever tip. A schematic diagram describing the experimental details has been provided in Figure 2(a). A tip made of silicon coated with 5 nm of titanium and 20 nm of iridium and spring constant 9N/m was used to write a two-dimensional array of square domains lithographically as shown in Figure 1(b). Each square domain was written by applying 40V dc on 8$\mu$m $\times$ 8$\mu$m area on the LVO surface. After writing the domain, the area was scanned in the contact mode while applying a small ac bias ($V_{ac} = 5V$) on the tip. An image of the phase for 4ML LVO-STO as a function of the position of the tip (the phase image) is shown in Figure 2(b). Similar phase image for 10ML LVO-STO was also observed. The phase image shows the square pattern written clearly while no substantial change in the topography was observed. In order to demonstrate the erasing capabilities, we applied a negative dc voltage i.e. -40V in the half of the central square in the pattern which erased half of the domain. In Figure 2(c) and 2(d), it is clearly seen that the half of the square in the center of the pattern is erased for 4ML and 10ML samples respectively. This mimicks ferroelctric-like switching with the 180$^\circ$C phase change with the application of the negative voltage\cite{Jyotsna}.

We have also performed local phase-voltage spectroscopy following similar protocol as in switching spectroscopy piezoresponse force microscopy (SS-PFM) in order to confirm the existence of local charging in the patterned region. In Figure 3(a) and Figure 3(c) we show the results on hysteretic switching of phase with the dc bias\cite{Sekhon} obtained at room temperature on two different thickesses, 4ML and 10ML respectively, of LVO on STO substrate. The Phase-voltage spectra show hysteresis as in ferroelectric samples. In Figures 3(b) and Figure 3(d), the amplitude response with dc bias for the same samples are shown, where the hysteretic spectra closesly resemble the "butterfly loops"\cite{Kalinin} usually seen in piezoelectric samples. The results hint on the presence of ferroelectric-like nature in LVO/STO systems.  However, ferroelectricity is not known to exist in such systems\cite{Sheets}. Therefore, it is rational to conclude that the nano-domains are primarily made of locally induced charges. The observed "butterfly loops" could originate from the strain developed locally due to local charge migration effects underneath the metallic tips.

\begin{figure}[t]
\centering
\includegraphics[width=0.70\textwidth]{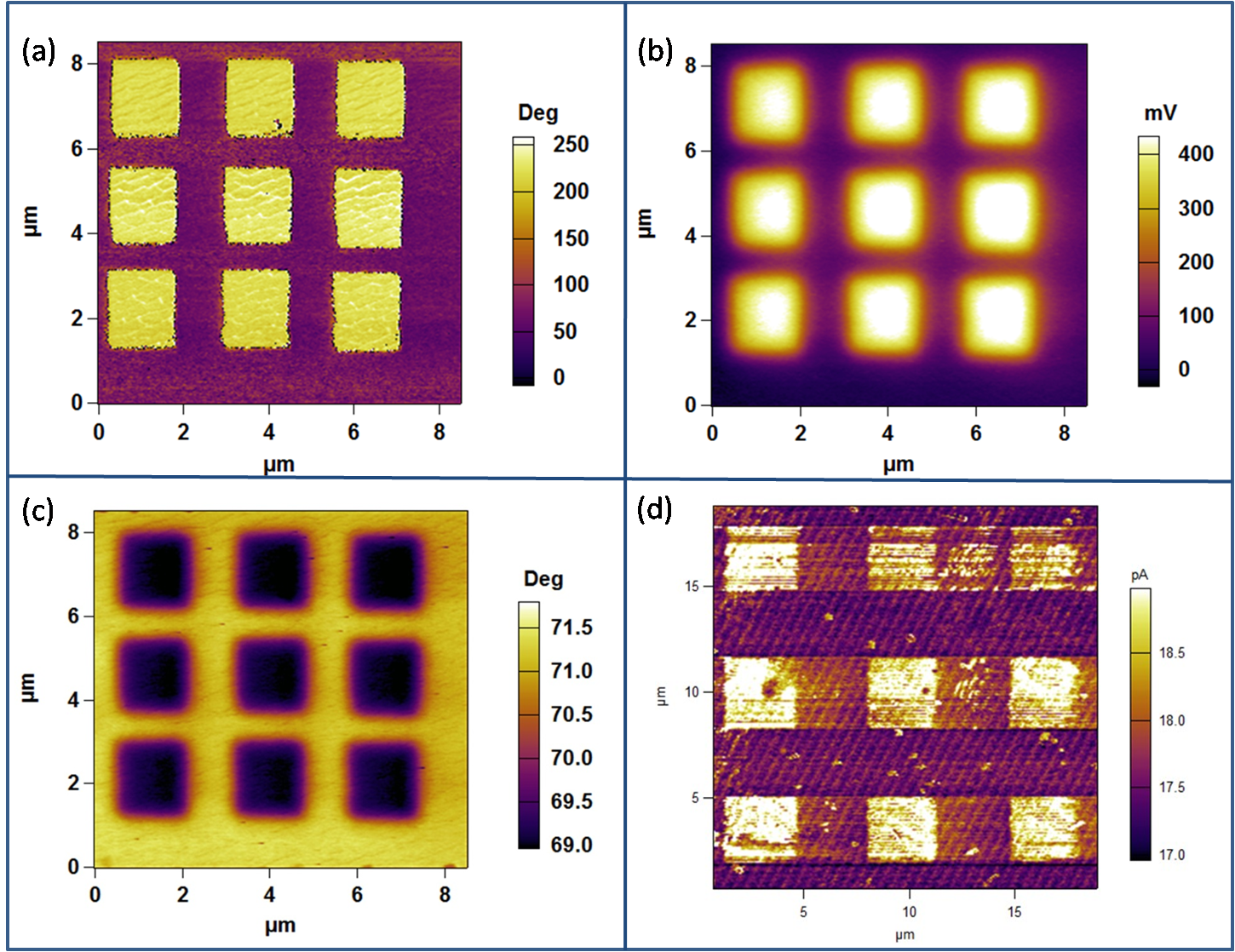}
\caption{(a) Phase image of the square patterned domains of the 4ML LVO on the STO substrate in the PFM mode (b) Potential map of the square patterned domains in the KPFM  (c) Phase image of the square patterned domains in the EFM mode (d) Conductivity variation of the domains. All above results are recorded at the room temperatures.}
	\label{Figure 4}
\end{figure}

\begin{figure}[t]
\centering
\includegraphics[width=0.70\textwidth]{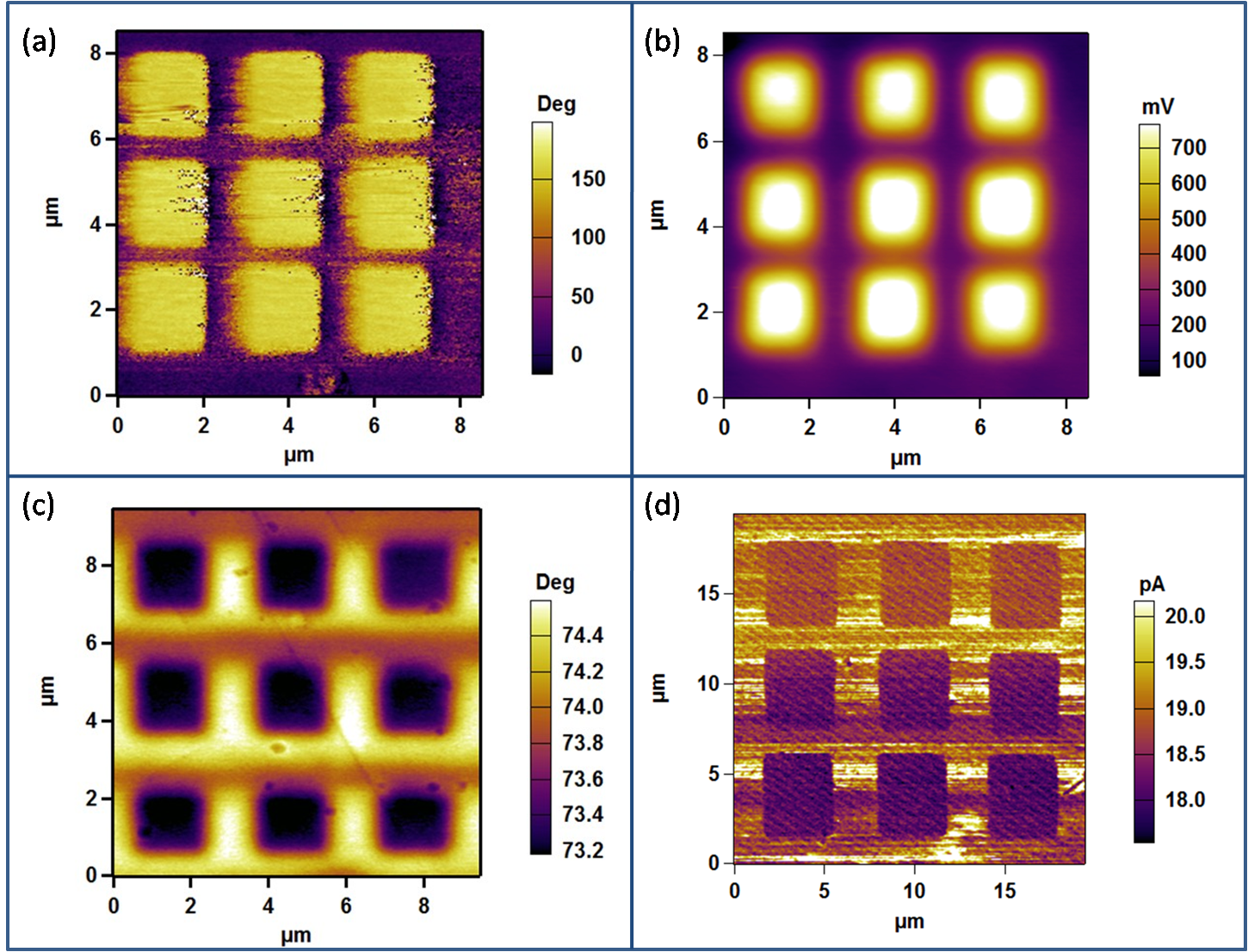}
\caption{(a) Phase image of the square patterned domains of the 10ML LVO on STO substrate in the PFM mode (b) Potential map of the square patterned domains in the KPFM  (c) Phase image of the square patterned domains in the EFM mode (d) Conductivity variation of the domains. All above results are recorded at the room temperatures.}
	\label{Figure 5}
\end{figure}

Further, in order to investigate the change in the physical properties of the patterned nanostructures, we have performed kelvin probe force microscopy (KPFM), electrostatic force microscopy (EFM) and conducting microscopy studies. Figure 4 highlights our findings of the nanofabricated square patterned artificial lattices in the 4ML LVO/STO 2DEG systems. Figure 4(a) shows the phase image of the square patterned domains written at room temperature in the PFM mode. Kelvin Probe Force Microscopy (KPFM) is primarily an AFM instrumentation tool to obtain the contact potential between the conducting AFM tip and the sample at high spatial resolution\cite{Nonn, Fuji, Melt}.  A potential change of 400mV is clearly seen Figure 4(b), where the bright regions on the pattern show higher potential and the dark regions show lower potential. Electrostatic force microscopy (EFM) probes the electrostatic force which arises due to the attraction or repulsion from surface charge. It maps locally charged domains on the sample surface, similar to as MFM plots the magnetic domains of the sample surface\cite{Kanto, Xu, Girard}. Figure 4(c) distinctly shows the charged domains in the phase image of the square patterned domains in the EFM mode. Thus, our EFM results can be used to study the spatial variation of surface charge carriers. And the charge carriers are trapped which are confined in the two dimensional electron gas at the interface of the LVO on STO. In figure 4(d), the conductivity variation\cite{Lei} of these square patterned domains with a change of 1-2pA is observed. This suggests that the domains in these artificially created lattices trap the charge carriers with perfectly coinciding potential modulation and conductivity mapping.

The comparative spectroscopic measurements were performed for the 10ML LVO on the STO substrate. Figure 5(a) shows the phase image of the square patterned domains written at room temperature in the PFM mode. It is observed that the phase image of the 10ML LVO/STO sample is not as sharp as the 4ML LVO/STO sample as seen in Figure 4(a). This can be attributed to the higher conductivity of 10ML LVO/STO sample with charge carrier density(n) as 1.09$\times10^{14} cm^{-2}$ as compared to $n=10^{11} cm^{-2}$ for 4ML LVO/STO sample. With higher charge carrier density, the charge carriers diffuse and result in a less distinct lithographically written nanostructures. A potential map of the square patterned domains is shown in Figure 5(b) which shows a distinct potential variation of 700mV. The charged domains in the phase image of the square patterned artificial lattice is observed in Figure 5(c). Similar to the 4ML LVO on STO, the electrons are trapped in the square patterned domains. The conductivity variation of the domains with a variation of 1-2pA is shown in Figure 5(d). Note that contrast to the 4ML LVO on STO substrate, the conductivity profile in the 10ML LVO on STO is found to be decreasing from the LVO surface to the square nanostructured domains due to the negative applied voltage. With the higher thickness of LVO deposited on STO substrate, the nanostructures are observed to show higher potential variation.

In conclusion, high quality nanofabricated artificial lattices hosting induced charges is achieved by AFM lithography. This yields an external potential landscape in the square geometry that acts as a lattice of potential wells to trap electron or holes. The spatial resolution of these structures is in the range of 1$\mu$m - 2 $\mu$m,  can be further improved to a few nanometeres using UHV STM. This approach allows us to independently tune the geometry, shape and inter-structure distances, which gives a unique interplay between the on-site interaction, nearest neighbor repulsive interactions and single particle hopping energy. These phenomenons pave a way for the dynamic control on the collective behavior and quantum phase transitions in such artificial lattices of solid state systems.

G.S. would like to acknowledge partial financial support from a research grant of a Ramanujan Fellowship awarded by the Department of Science and Technology (DST), Govt. of India under grant number SR/S2/RJN-99/2011 and a research grant from DST-Nanomission under grant number
SR/NM/NS-1249/2013. MB thanks DST-Nanomission for the JRF position. SC acknowledges the financial support of DST Nano Mission project number (SR/NM/NS-1007/2015). NW and SC acknowledge the financial support of Funding Program for World-Leading Innovative R \& D on Science and Technology (FIRST) of the Japan Society for the Promotion of Science (JSPS) initiated by the Council for Science and Technology Policy, by JSPS Grants-in Aid for Scientific Research, No. 24226002.

%\section*{\bfseries{REFERENCES}}

\end{document}